\documentclass[twoside]{article}
\usepackage{fleqn,espcrc2}
\usepackage{epsfig}
\def\lsim{\raise0.3ex\hbox{$<$\kern-0.75em\raise-1.1ex\hbox{$\sim$}}}
\def\gsim{\raise0.3ex\hbox{$>$\kern-0.75em\raise-1.1ex\hbox{$\sim$}}}
\usepackage{graphicx}
\usepackage[figuresright]{rotating}

\newcommand{\AmS}{{\protect\the\textfont2
  A\kern-.1667em\lower.5ex\hbox{M}\kern-.125emS}}

\title{
Flavor and Quark Mass Dependence of QCD Thermodynamics}
\author{F. Karsch, E. Laermann, A. Peikert, Ch. Schmidt and S. Stickan 
\thanks{The work has been supported by the TMR network ERBFMRX-CT-970122
and the DFG under grant Ka 1198/4-1 .}
\\
\vskip 6pt
Fakult\"at f\"ur Physik, Universit\"at Bielefeld, D-33615 Bielefeld,
Germany
}

\begin{document}

\begin{abstract}
We calculate the transition temperature in 2 and 3-flavor QCD
using improved gauge and staggered fermion actions on lattices with
temporal extent $N_\tau=4$. We find $T_c=173\; (8)\; $MeV and $154\;
(8)\; $MeV for $n_f=2$ and 3, respectively. In the case of 3-flavor QCD 
we present evidence that the chiral critical point,  {\it i.e.} the second 
order endpoint of the line of first order chiral phase transitions, 
belongs to the universality class of the 3d Ising model. 
\vspace{1pc}
\end{abstract}

\maketitle

\section{Introduction}
Determining the properties of strongly interacting matter at high
temperature, the temperature at which the transition to the quark-gluon
plasma phase takes place as well as other critical parameters, 
is one of the basic goals in lattice studies of finite-T QCD. 
Numerical calculations which have
been performed during the last years have shown that the details of
the transition strongly depend on the number of quark flavors ($n_f$) as well
as the value of e.g. the pseudo-scalar meson mass ($m_{\rm PS}$) which is 
controlled by the bare quark masses ($m_q$). Furthermore, it became evident 
that the moderate values of the lattice spacing ($a \sim 0.25$fm) used
in finite-T calculations with dynamical fermions lead to sizeable cut-off 
effects;
different discretization schemes used in the fermion sector, e.g.
the standard staggered and Wilson fermion formulation, lead to significantly 
different results for $T_c$ \cite{Kar99}. Calculations with improved 
actions thus seem to be mandatory to obtain reliable quantitative results for 
$T_c$ and its dependence on $n_f$ as well as $m_{\rm PS}$.

We will report here on results for $n_f =2$ and 3 obtained with 
a Symanzik improved gauge action and a staggered fermion action with improved
rotational invariance,
{\it i.~e.} the p4-action with fat 1-link terms. This action 
has been used previously by us in 
studies of the flavor dependence of the QCD thermodynamics  
\cite{Kar00}. Further details on the 
action and the algorithm used for our simulations are given in \cite{Kar00}. 
For a discussion of recent results with improved Wilson fermions see 
Ref.~\cite{Ali00}.

\section{Flavor and quark mass dependence of $T_c$}

The transition temperature in QCD with dynamical fermions has been
found to be significantly smaller than in the pure gauge sector. This
is in accordance with intuitive pictures of the phase transition based
for instance on the thermodynamics of bag or percolation models. 
With decreasing $m_q$ hadrons become lighter and it becomes easier 
to build up a sufficiently high particle density that can trigger a 
phase transition.  For the same reason such models also suggest that $T_c$
decreases when $n_f$ and in turn the number of light pseudo-scalar 
mesons increases. This general picture is confirmed by the lattice results 
presented in the following.

\begin{figure*}
\begin{center}
\epsfig{file=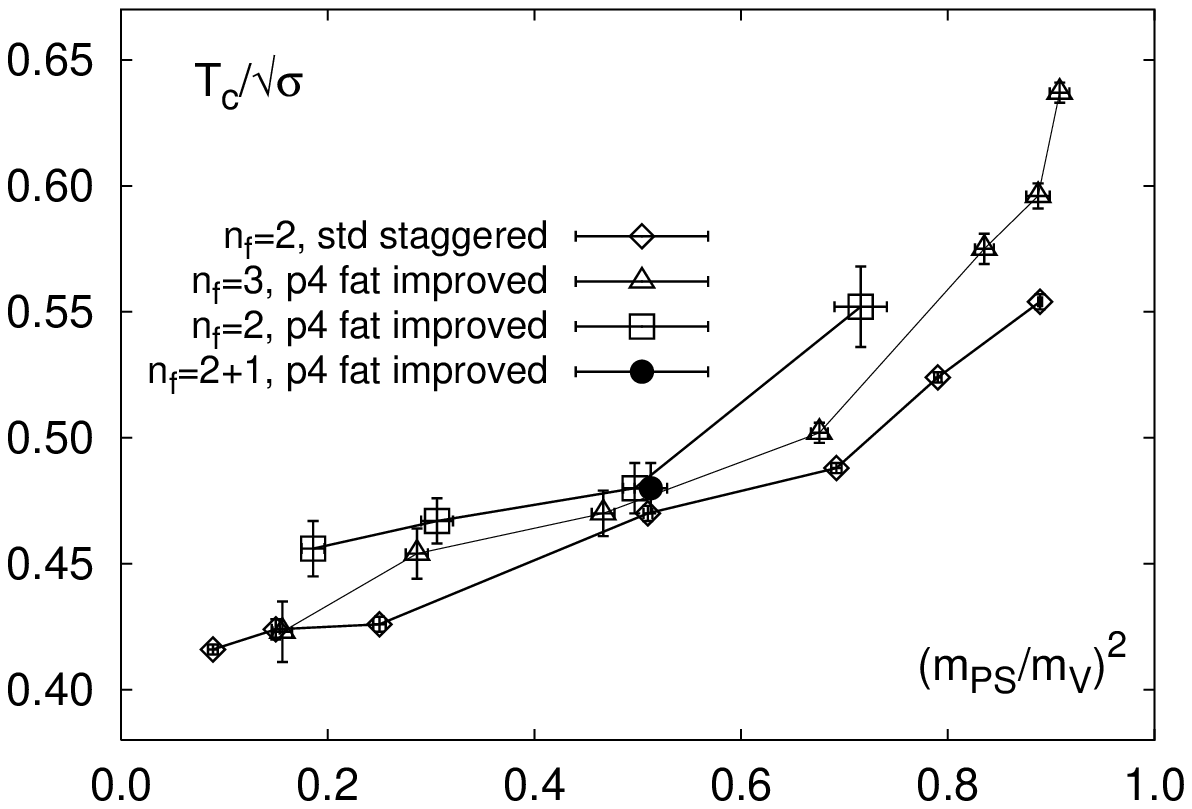,width=74mm}
\epsfig{file=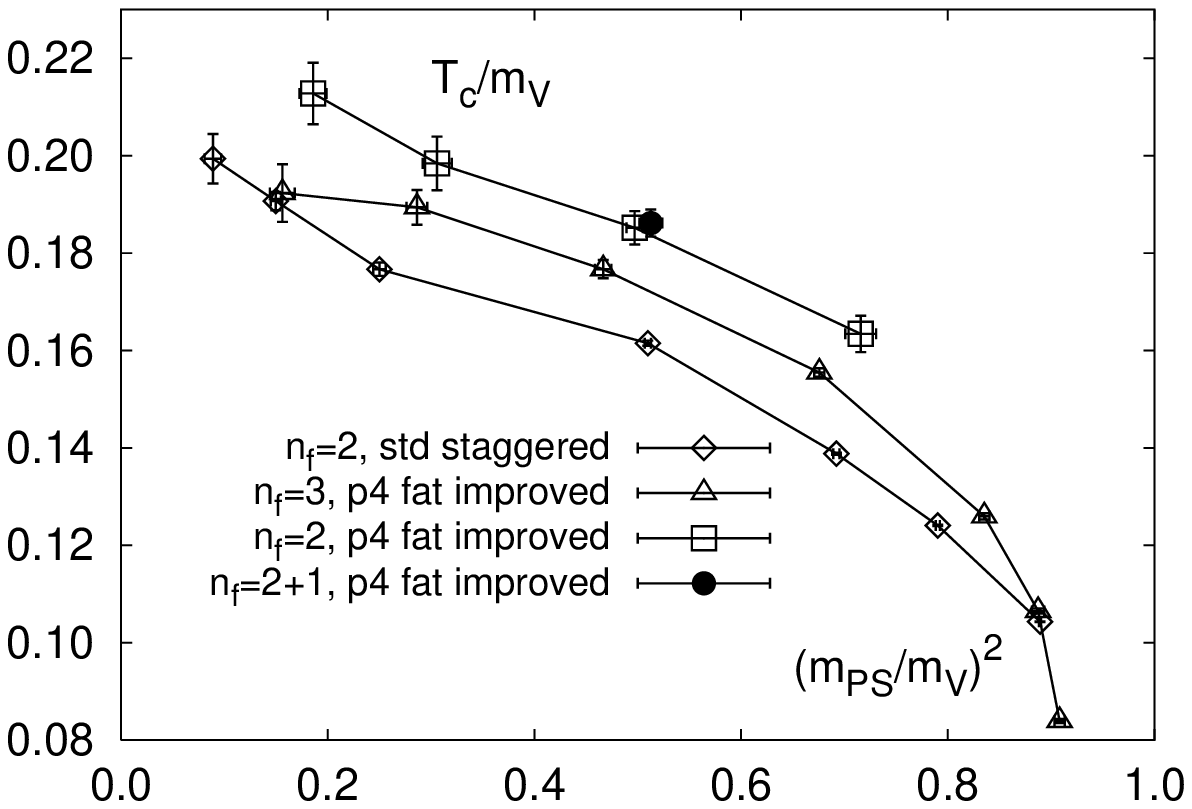, width=74mm}
\end{center}
\vspace*{-0.7cm}
\caption{The critical temperature in units of $\sqrt{\sigma}$
(left) and the vector meson mass (right) versus
$(m_{\rm PS}/m_{\rm V})^2$. Shown are results for 2, (2+1) and 3-flavor
QCD obtained from calculations with the p4 action
on lattices with temporal extent $N_\tau=4$. For $n_f=2$ we also show
results obtained by using unimproved gauge and staggered fermion
actions.}
\label{fig:crit_temp}
\end{figure*}

We have calculated the pseudo-critical couplings, $\beta_c(m_q)$, 
for the transition to the QCD plasma phase on lattices of size 
$16^3\times 4$.  Calculations have been performed for 2 and 3-flavor 
QCD in a wide range of quark
masses, $0.025 \le m_q \le 1.0$. The smallest quark mass value 
corresponds to $m_{\rm PS} \simeq 350$~MeV. 

In order to set a scale
for the transition temperature we have performed calculations of the light
meson spectrum and the string tension\footnote{We note that the heavy quark
potential is no longer strictly confining in the presence of dynamical
quarks. The definition of the string tension is based on potentials
extracted from Wilson loops which have been found to lead to a linear 
rising potential at least up to distances $r\sim 2$fm.}
at $\beta_c(m_q)$ on lattices of size $16^4$. The results of this
calculation are shown in Fig.~\ref{fig:crit_temp}.

We note that $T_c/\sqrt{\sigma}$ and $T_c/m_V$ do show a consistent
flavor dependence of $T_c$. In a wide range of quark mass values $T_c$ 
in 3-flavor QCD is about 10\% smaller than for $n_f=2$. 
The dependence on $m_q$ is, however, quite different when using the vector 
meson mass rather than the string tension to set the scale for $T_c$.  
While $T_c/\sqrt{\sigma}$ does show the expected rise with increasing
$m_q$ the contrary is the case for $T_c/m_V$. Of
course, this does not come as a surprise. The vector meson mass used in
Fig.~\ref{fig:crit_temp} to set the scale is itself strongly quark mass
dependent, $m_V = m_\rho +c\; m_q$. Its mass thus is significantly larger
than the vector meson mass, $m_\rho$, in the chiral limit.
We stress this well
known fact here because it makes evident that one has to be careful
when discussing the dependence of $T_c$ on parameters of the QCD
Lagrangian, e.g. $n_f$ and $m_q$. One has
to make sure that the observable used to set the scale for $T_c$
itself is not or at most only weakly dependent on the external parameters.

\begin{figure}
\begin{center}
\epsfig{file=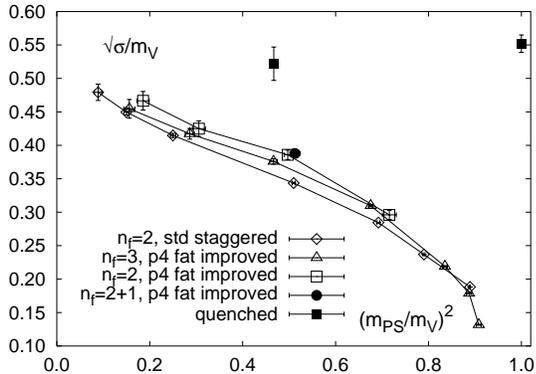,width=75mm}
\end{center}
\vspace{-0.7cm}
\caption{$\sqrt{\sigma}/m_V$ versus $(m_{\rm PS}/m_V)^2$ for 
2, 2+1 and 3-flavor
QCD obtained with the p4-action.  Also shown are results obtained
with unimproved staggered fermions.
The filled black squares show results from quenched and partially
quenched calculations with a sea quark mass of $m_q=0.1$.}
\label{fig:sigma_rho}
\end{figure}

In Fig.~\ref{fig:sigma_rho} we show results for the ratio 
$\sqrt{\sigma}/m_V$ calculated on $16^4$ lattices for various values
of the quark mass in 2, (2+1) and 3-flavor QCD. Also shown there is the result
obtained in quenched QCD as well as from a partially quenched calculation
which we have performed on gauge field configurations generated with a dynamical
(sea) quark mass of $m_q=0.1$. The fact that this ratio shows only little 
flavor dependence and that the (partially) quenched limit is in reasonable
agreement with experiment and phenomenology\footnote{Systematic deviations
are visible but are on the 10\% level.} 
indicates that the corresponding partially quenched
observables are suitable also for setting the scale in the presence of
dynamical quarks.

\subsection{$\bf T_c$ and thermodynamics of (2+1)-Flavor QCD}
Physically most relevant is an analysis of the thermodynamics of
QCD with 2 light ($m_u\simeq m_d\simeq 0$) and a heavier (strange) 
quark with $m_s\simeq T_c$. Calculations of bulk thermodynamic
observables (pressure, energy density) suggest that the explicit 
quark mass dependence is small above $T_c$. In particular, a recent
calculation of the pressure with $m_s/T=1$ and 
$m_{u,d}/T=0.4$ has shown that deviations from the case with three 
degenerate quarks are only about $10\%$ \cite{Kar00}. 
This difference would be even smaller when $m_s$ rather than
$m_s/T$ would have been kept fixed in the simulation. 
One thus can expect that already for $T\gsim 2\; T_c$ 
bulk thermodynamics of the plasma phase will be well described by
the equation of state of 3 degenerate, nearly massless flavors.
This is different close to the transition which in the (2+1)-flavor
case presumably is just a rapid crossover. Although this analysis
has not yet been performed with a sufficiently light
light quark sector the result for the pseudo-critical temperature
shown in Fig.~1 suggests that $T_c$ is close to that of 2-flavor QCD. 

\subsection{Cut-off effects}
In Fig.~1 we also show results for $T_c$ obtained with the 
standard, unimproved gauge and staggered fermion actions. We note that
on lattices with temporal extent $N_\tau=4$ this action leads 
to systematically smaller values for $T_c/\sqrt{\sigma}$ as well as 
$T_c/m_V$. This provides an estimate for systematic cut-off effects,
which for a wide range of quark mass values seem to be about 
$10\%$. Such an estimate also is consistent  
with the observation that for large values of $N_\tau$ 
calculations with unimproved fermion actions do lead to larger critical 
temperatures. Statistical errors are, however, large in this case.
%on $T_c$ do increase too. 
We thus expect that 
our current analysis with improved gauge and staggered fermion actions
is affected with systematic cut-off effects which are significantly below
the $10\%$ level. However, this clearly has to be checked with simulations
on larger lattices.

\section{$\bf T_c$ in the chiral limit}

For $n_f=3$
the chiral phase transition is known to be first order \cite{old3}
whereas it is most likely a continuous transition for $n_f=2$.
This also implies that the dependence of the (pseudo)-critical
temperature on the quark mass will differ in both cases.
Asymptotically, {\it i.e.}
to leading order in the quark mass one expects to find,

\begin{equation} 
T_c (m_{\rm PS}) -T_c(0) \sim \cases{ m_{\rm PS}^{2/\beta\delta} &, $n_f=2$ \cr
 m_{\rm PS}^2 &, $n_f \ge 3$}\quad ,
\label{tcscaling}
\end{equation}
with $1/\beta\delta=0.55$ if the 2-flavor transition indeed belongs to the
universality class of 3d, O(4) symmetric spin models. Our estimates of
$T_c$ in the chiral limit are based on the data shown in Fig.~1. For
$n_f=3$ we have extrapolated $T_c/\sqrt{\sigma}$ and $T_c/m_V$
using an ansatz quadratic in $m_{\rm PS}/m_V$. In addition we have
extrapolated $m_V$ calculated at $\beta_c(m_q)$ for $m_q=0.025$ and 0.05
linearly in $m_q$ to the critical point in the chiral limit,
$\beta_c(0) = 3.258(4)$. The extrapolation to the chiral limit is
less straightforward in the case of $n_f=2$. The data shown in Fig.~1 indicate
that the quark mass dependence for $n_f=2$ and 3 is quite similar. This
suggests that sub-leading corrections, quadratic in $m_{\rm PS}/ m_V$,
should be taken into account in addition to the leading behavior
expected from universality. In our extrapolations for $n_f=2$ we thus
also add a term quadratic in $m_{\rm PS}/m_V$ to the leading term given
in Eq.~1.  

\begin{figure*}[t]
\vspace*{-1.7cm}
\begin{center}
\hspace*{-0.8cm}
\epsfig{file=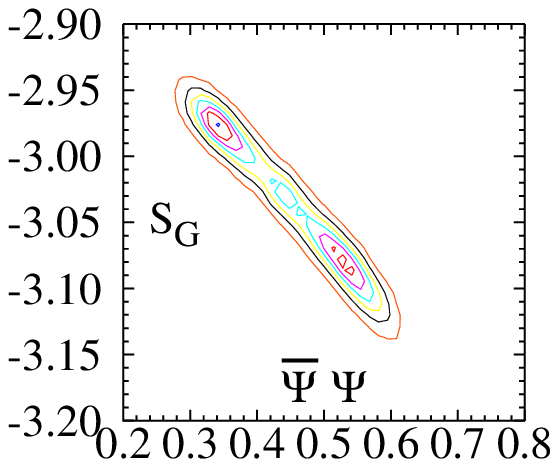,width=66mm}\hspace*{-2.7cm}
\epsfig{file=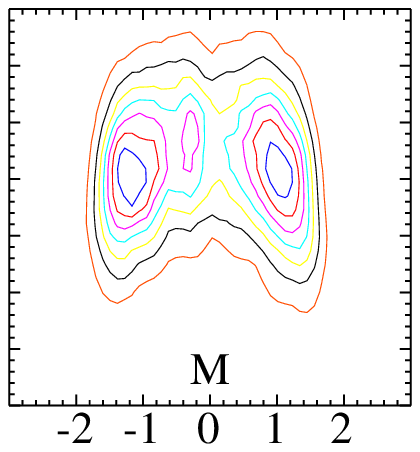,width=66mm}\hspace*{-3.5cm}
\epsfig{file=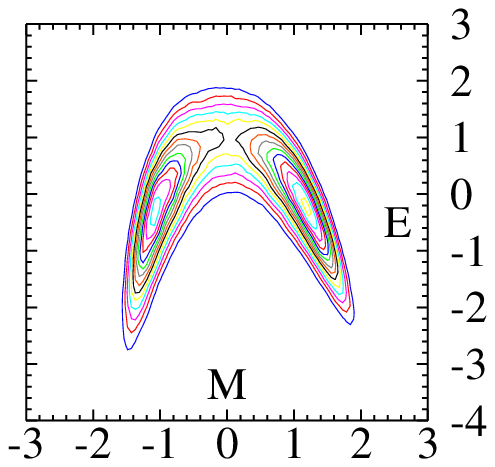,width=66mm}
\end{center}
\vspace*{-2.5cm}
\label{fig:histo_EM}
\caption{Contour plot for the joint probability distribution of $S_G$
and $\bar{\psi}\psi$ (left) as well as $E$ and
$M$ (middle) for 3-flavor QCD and the 3d, 3-state Potts model (right).
The QCD contour plots are based on calculations
performed on a $16^3\times 4$ lattice at $\beta=5.1499$ with
$m_q=0.035$. These parameters are in agreement with a previous
estimate for the location of the chiral critical point \cite{Aoki99}. 
The mixing parameters have been fixed to $r=0.55$ and $s=0$. Parameters 
for the Potts model are given in Ref.~7.
}
\end{figure*}

\noindent
From these fits we estimate
\begin{equation}
{T_c \over m_\rho} = \cases{ 0.225 \pm 0.010 &, $n_f=2$ \cr
0.20 \pm 0.01 &, $n_f = 3$}\quad ,
\end{equation}
which corresponds to $T_c=(173 \pm 8)~{\rm MeV}$ and
$(154 \pm 8)~{\rm MeV}$ for $n_f=2$ and 3, respectively.
For $n_f=2$ this agrees well with results obtained in a calculation with
improved Wilson fermions \cite{Ali00}. We stress, however, that
the errors given here as well as in Ref.~3 are statistical only.
Systematic errors resulting
from remaining cut-off effects and from the ansatz used for extrapolating
to the chiral limit are expected to be of similar magnitude. In order
to control these errors calculations on lattices with larger
temporal extent are still needed. 

\section{Universality at the chiral critical point of 3-flavor QCD}
The first order chiral transition of 3-flavor QCD 
will persist to be first order for $m_q > 0$ up to a
critical value of the quark mass. At this {\it chiral critical
point} the transition will be second order and it has been conjectured
that it belongs to the universality class of the
3d Ising model \cite{Gav94}. From a simulation with standard, {\it i.e.}
unimproved, gauge and staggered fermion actions we find support for this
conjecture.

The analysis of the critical behavior at the $2^{\rm nd}$ order endpoint of 
a line of $1^{\rm st}$ order phase transitions requires the correct 
identification of
energy-like and ordering-field (magnetic) directions \cite{higgs,Stickan}.
The ordering field at the critical point
can be constructed from a linear combination of the gluonic action $S_G$
and the chiral condensate $\bar{\psi}\psi$,
\begin{equation}
E = S_G + r\; \bar{\psi}\psi \quad,\quad
M = \bar{\psi}\psi + s\; S_G \quad.
\label{em}
\end{equation}
Here the mixing parameter $r$ is determined from the $m_q$-dependence
of the line of first order transitions, $r=({\rm d} \beta /{\rm d} m_q
)^{-1}_{\rm endpoint}$ and $s$ by demanding $\langle E\cdot M\rangle =
0$.
In fact, unlike for energy-like observables, e.g. critical amplitudes
that involve the thermal exponent $y_t$, the universal properties of
observables related to $M$ do not depend on the correct choice of
the mixing parameter $s$ as long as the magnetic exponent $y_h$
is larger than $y_t$. The joint probability distribution for $E$ and $M$
characterizes the symmetry at the critical endpoint and
its universality class. Contour plots for the normalized $E$-$M$
distributions at the critical endpoints of 3-flavor QCD and the
3d, 3-state Potts model as well as the corresponding plot
for the $S_G$-$(\bar{\psi}\psi)$ distribution are given in Fig.~3.
This shows that also in the QCD case a proper definition of energy-like
and ordering-field operators is needed to reveal the symmetry
properties of the chiral critical point. The joint distributions of $E$
and $M$ suggest the universal structure of the
$E$-$M$ probability distribution of the 3-d Ising model \cite{higgs},
although it is apparent that an analysis of 3-flavor QCD on larger
lattices
is needed to clearly see the ``two wings''
characteristic for the 3d Ising distribution.

\end{document}